\documentclass{article}
\usepackage{graphics}
\usepackage{comment}
\usepackage{amsmath}
\begin{document}

\def\tr{ {\rm{Tr \,}}}
\def\diag{ {\rm{diag \,}}}

\def\supp{ {\rm{supp \,}}}
\def\dim{ {\rm{dim \,}}}
\def\oti{{\otimes}}
\def\bra#1{{\langle #1 |  }}
\def\lb{ \left[ }
\def\rb{ \right]  }
\def\tilde{\widetilde}
\def\bar{\overline}
\def\*{\star}

\def\({\left(}		 		 \def\BL{\Bigr(}
\def\){\right)}		 		 \def\BR{\Bigr)}
		 \def\BBL{\lb}
		 \def\BBR{\rb}
%

\def\bb{{\bar{b} }}
\def\ab{{\bar{a} }}
\def\zb{{\bar{z} }}
\def\zbar{{\bar{z} }}
\def\frac#1#2{{#1 \over #2}}
\def\inv#1{{1 \over #1}}
\def\half{{1 \over 2}}
\def\d{\partial}
\def\der#1{{\partial \over \partial #1}}
\def\dd#1#2{{\partial #1 \over \partial #2}}
\def\vev#1{\langle #1 \rangle}
\def\ket#1{ | #1 \rangle}
\def\proj#1{ | #1 \rangle \langle #1 |}
\def\rvac{\hbox{$\vert 0\rangle$}}
\def\lvac{\hbox{$\langle 0 \vert $}}
\def\2pi{\hbox{$2\pi i$}}
\def\e#1{{\rm e}^{^{\textstyle #1}}}
\def\grad#1{\,\nabla\!_{{#1}}\,}
\def\dsl{\raise.15ex\hbox{/}\kern-.57em\partial}
\def\Dsl{\,\raise.15ex\hbox{/}\mkern-.13.5mu D}
\def\b#1{\mathbf{#1}}
%
%
\def\th{\theta}		 		 \def\Th{\Theta}
\def\ga{\gamma}		 		 \def\Ga{\Gamma}
\def\be{\beta}
\def\al{\alpha}
\def\ep{\epsilon}
\def\vep{\varepsilon}
\def\la{\lambda}		 \def\La{\Lambda}
\def\de{\delta}		 		 \def\De{\Delta}
\def\om{\omega}		 		 \def\Om{\Omega}
\def\sig{\sigma}		 \def\Sig{\Sigma}
\def\vphi{\varphi}
%
%
\def\CA{{\cal A}}		 \def\CB{{\cal B}}		 
\def\CC{{\cal C}}
\def\CD{{\cal D}}		 \def\CE{{\cal E}}		 
\def\CF{{\cal F}}
\def\CG{{\cal G}}		 \def\CH{{\cal H}}		 
\def\CI{{\cal J}}
\def\CJ{{\cal J}}		 \def\CK{{\cal K}}		 
\def\CL{{\cal L}}

\def\CM{{\cal M}}		 \def\CN{{\cal N}}		 
\def\CO{{\cal O}}
\def\CP{{\cal P}}		 \def\CQ{{\cal Q}}		 
\def\CR{{\cal R}}
\def\CS{{\cal S}}		 \def\CT{{\cal T}}		 
\def\CU{{\cal U}}
\def\CV{{\cal V}}		 \def\CW{{\cal W}}		 
\def\CX{{\cal X}}
\def\CY{{\cal Y}}		 \def\CZ{{\cal Z}}
\newcommand{\qed}{\rule{7pt}{7pt}}
\def\E{{\mathbf{E} }}
\def\1{{\mathbf{1} }}
\def\rvac{\hbox{$\vert 0\rangle$}}
\def\lvac{\hbox{$\langle 0 \vert $}}
\def\comm#1#2{ \BBL\ #1\ ,\ #2 \BBR }
\def\2pi{\hbox{$2\pi i$}}
\def\e#1{{\rm e}^{^{\textstyle #1}}}
\def\grad#1{\,\nabla\!_{{#1}}\,}
\def\dsl{\raise.15ex\hbox{/}\kern-.57em\partial}
\def\Dsl{\,\raise.15ex\hbox{/}\mkern-.13.5mu D}
\def\beq{\begin {equation}}
\def\eeq{\end {equation}}
\def\to{\rightarrow}

\title{Quantum Teleportation and Jungian Psychology}

\author{
Igor Devetak\footnote{The author is with Systematica Investments, 1201 Geneva, Switzerland. Email address: {\tt igordevetak@gmail.com}}\\
} 

\date{\today} 
\maketitle

\begin{abstract}
We propose that the Jungian psychological type of an individual is naturally modelled as a quantum state: a maximally entangled two-qubit state, one of whose qubits is undergoing quantum teleportation.   
\end{abstract}

\section{Introduction}
A century ago Carl Gustav Jung, father of analytical psychology, introduced a rigorous theory of psychological types \cite{jung6}. According to Jung, every individual belongs to a particular psychological type. This is as definite a property of humans as their blood type or sex chromosomes, but less obvious how to measure.  The only shortcoming of Jung's theory is that it is incomplete --  much like real numbers are incomplete without imaginary ones. Because Jung's model could not fully account for all of the traits of an individual, rather than extending the model, his successors engaged in amending and reinterpreting it in a way that unfortunately marred some of its mathematical properties and elegance. 

Here we go back to  Jung's original work and augment it with a few other important concepts also introduced by Jung himself. At this point the model becomes sufficiently rich to mandate a proper mathematical representation. It turns out that a natural one is found in the formalism of quantum mechanics, in particular the quantum teleportation protocol \cite{bennett}.  

\section{Classical Jungian theory of types}
\subsection{The four elementary cognitive functions}
\begin{figure}
\centerline{ {\scalebox{.35}{\includegraphics{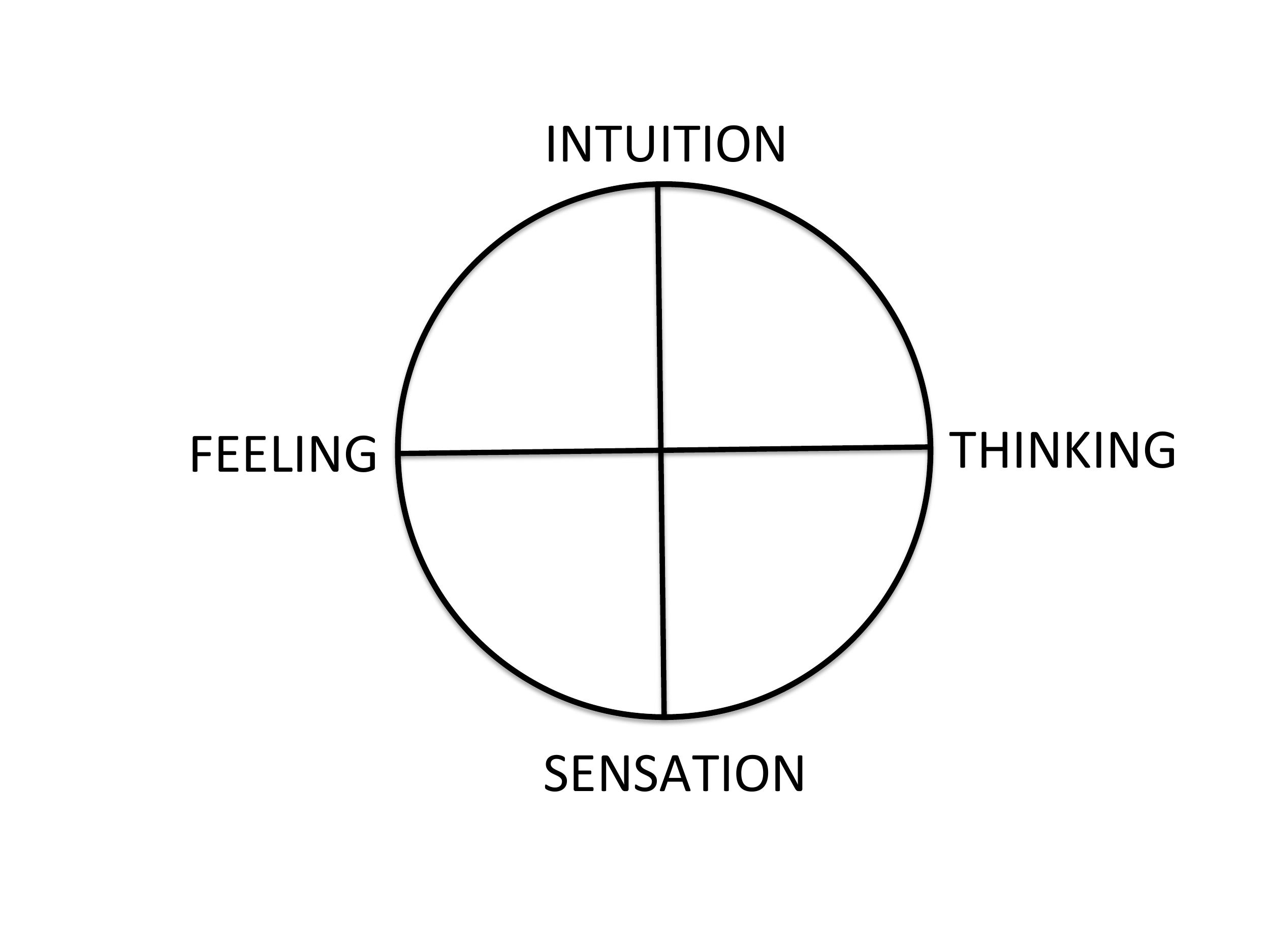}}}}
\caption{The circle of the four elementary cognitive functions} 
\label{fig:4}
\end{figure}

Jung defined the four elementary cognitive functions: intuition (N), sensation (S), thinking (T) and feeling (F). They are categorized into \emph{irrational} and \emph{rational}. 
\begin{enumerate}
\item The opposing irrational functions, intuition (N) and sensation (S), are neutral and have to do with perception. Sensation takes in raw information passively, as is, and simply experiences it. Intuition involves a fully active, collaborative approach in which we seek out the information, engage with it, interpret it.
\item The opposing rational functions, thinking (T) and feeling (F), have to do with judgment. They operate in the dual setting of true/false, like/dislike, etc. Thinking is judging based on objective logic and hard facts. Feeling is judging based on "soft" subjective considerations of an aesthetic or emotional nature.
\end{enumerate}
The elementary psychological functions are usually represented on a circle as the four cardinal directions (Fig. \ref{fig:4}). Thinking and feeling are diametric opposites, as are intuition and sensation. Rational and irrational functions should not be viewed as opposite but complementary. For instance the first quadrant of the circle corresponds to combining thinking and intuition, the second to combining intuition and feeling, etc.

\subsection{The eight composite cognitive functions}
In addition to irrationality (N vs. S) and rationality (T vs. F) there is a third dimension, namely \emph{attitude}. Jung defined a person to be extroverted (e) if naturally more interested in the outer, physical world and introverted (i) if naturally more interested in the inner world. Correspondingly, each of the four elementary functions comes in two flavours, extroverted and introverted, resulting in a total of eight composite cognitive functions: Ne, Ni, Se, Si, Te, Ti, Fe and Fi. We elaborate on their psychological meaning in Appendix \ref{A}.

\begin{figure}
\centerline{ {\scalebox{.4}{\includegraphics{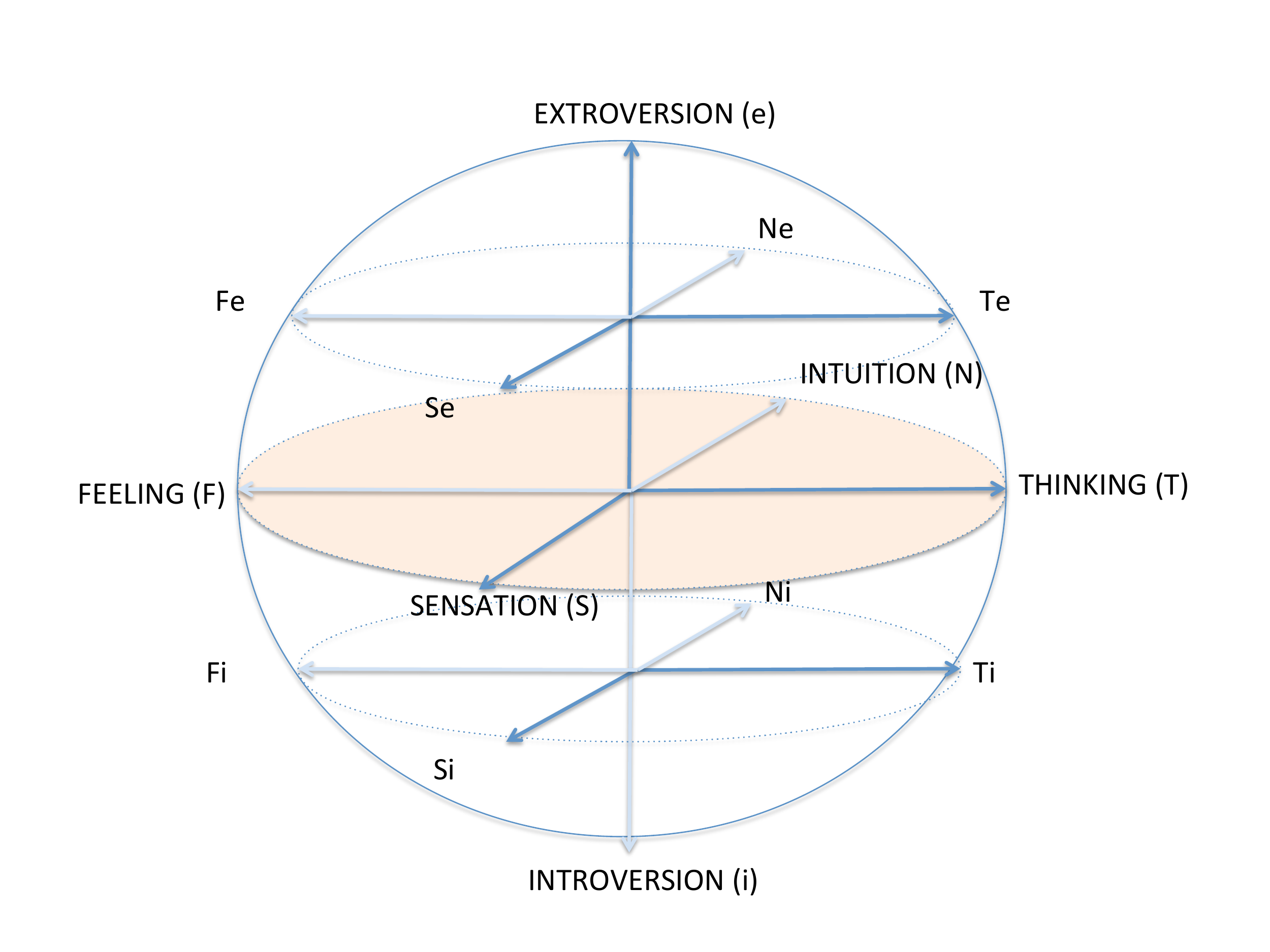}}}}
\caption{The sphere of the eight composite functions}
\label{fig:8}
\end{figure}
 
We can represent the eight composite psychological functions as points on a sphere (Fig. \ref{fig:8}). The equator is our circle with the four elementary psychological functions from Fig. \ref{fig:4}. The equatorial plane contains two orthogonal axes: the F-T axis and the S-N axis. The north and south poles represent extroversion (e) and introversion (i), respectively, thus defining the third e-i axis orthogonal to the previous two.
 
We place the Te composite function in the middle of the arc connecting the T and e points, the Si function on the arc connecting S and i, etc.   The eight composite functions now form four antipodal pairs: Ne-Si, Se-Ni, Te-Fi and Fe-Ti.

\subsection{Psychological types}
Psychological types live on the sphere of the eight composite functions. Jung's theory of types can be expressed by the following two postulates. For a given individual:
\begin{enumerate}
\item The conscious mind has a preferred point on the sphere of composite psychological functions.
\item The unconscious mind also has a preferred point: the point antipodal to the conscious one.
\end{enumerate}
We have snuck in a new duality here, namely that between the conscious and unconscious mind. 

\begin{figure}
\centerline{ {\scalebox{.40}{\includegraphics{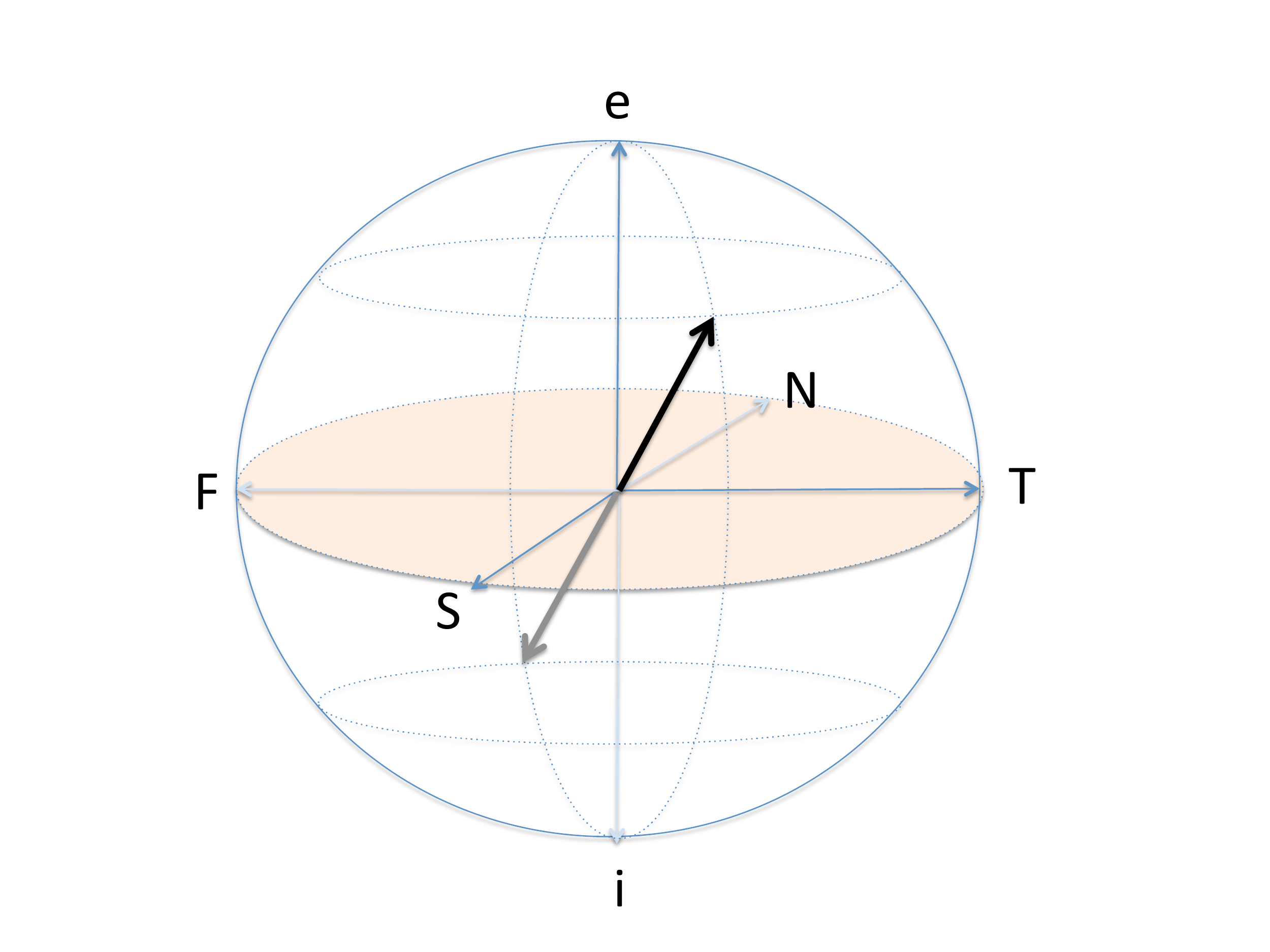}}}}
\caption{The conscious and unconscious psychological type} 
\label{fig:T}
\end{figure}

An example of the psychological type of a person is shown in Fig. \ref{fig:T}. The black vector represents someone who is extroverted (northern hemisphere) and has thinking and sensation as preferred functions. These are the functions that he uses most naturally with his conscious mind. Unconsciously however he is introverted (southern hemisphere) and prefers the feeling and intuition functions (grey vector). The unconscious perfectly balances the conscious -- this was Jung's brilliant insight. 

In practice it is possible to ascertain, based on  behaviour and personality traits, which octant of the sphere the conscious vector lies in (and thus the unconscious vector is in the octant antipodal to it). This "octant approximation" gives rise to eight distinct types defined in Table \ref{tab:types}. As it is often hard to distinguish between conscious and unconscious traits, there is a natural equivalence relation of interchanging the conscious and unconscious vectors.  This collapses the eight types into four equivalence classes, which we label  by the numbers 0--3. For example, type 1e lies in the octant defined by extroversion (e), feeling (F) and sensation (S), while 1i is antipodal to it: introversion (i), thinking (T) and intuition (N).

\begin{table}[]
\centering
\begin{tabular}{|c|c|c|}
\hline
Type & Conscious Functions & Unconscious Functions \\ \hline
0i   & Ti+Si               & Fe+Ne                 \\ \hline
0e   & Fe+Ne               & Ti+Si                 \\ \hline
1i   & Ti+Ni               & Fe+Se                 \\ \hline
1e   & Fe+Se               & Ti+Ni                 \\ \hline
2i   & Fi+Si               & Te+Ne                 \\ \hline
2e   & Te+Ne               & Fi+Si                 \\ \hline
3i   & Fi+Ni               & Te+Se                 \\ \hline
3e   & Te+Se               & Fi+Ni                 \\ \hline
\end{tabular}
\caption{The eight distinct psychological types of the "octant approximation"} 
\label{tab:types}
\end{table}

\section{The psychological type as a maximally entangled state}

In the previous section we have characterized the psychological type of a person as an ordered pair of antipodal points on a sphere. We will now relate this to a quantum state.

Consider a generic qubit state
$$
\ket{\psi} = \sin(\theta/2) \ket{0} + \cos(\theta/2)e^{i \varphi} \ket{1}
$$ 
where $\{\ket{0}, \ket{1}\}$ is a preferred orthonormal basis (a.k.a. computational basis).
\begin{figure}
\centerline{ {\scalebox{.4}{\includegraphics{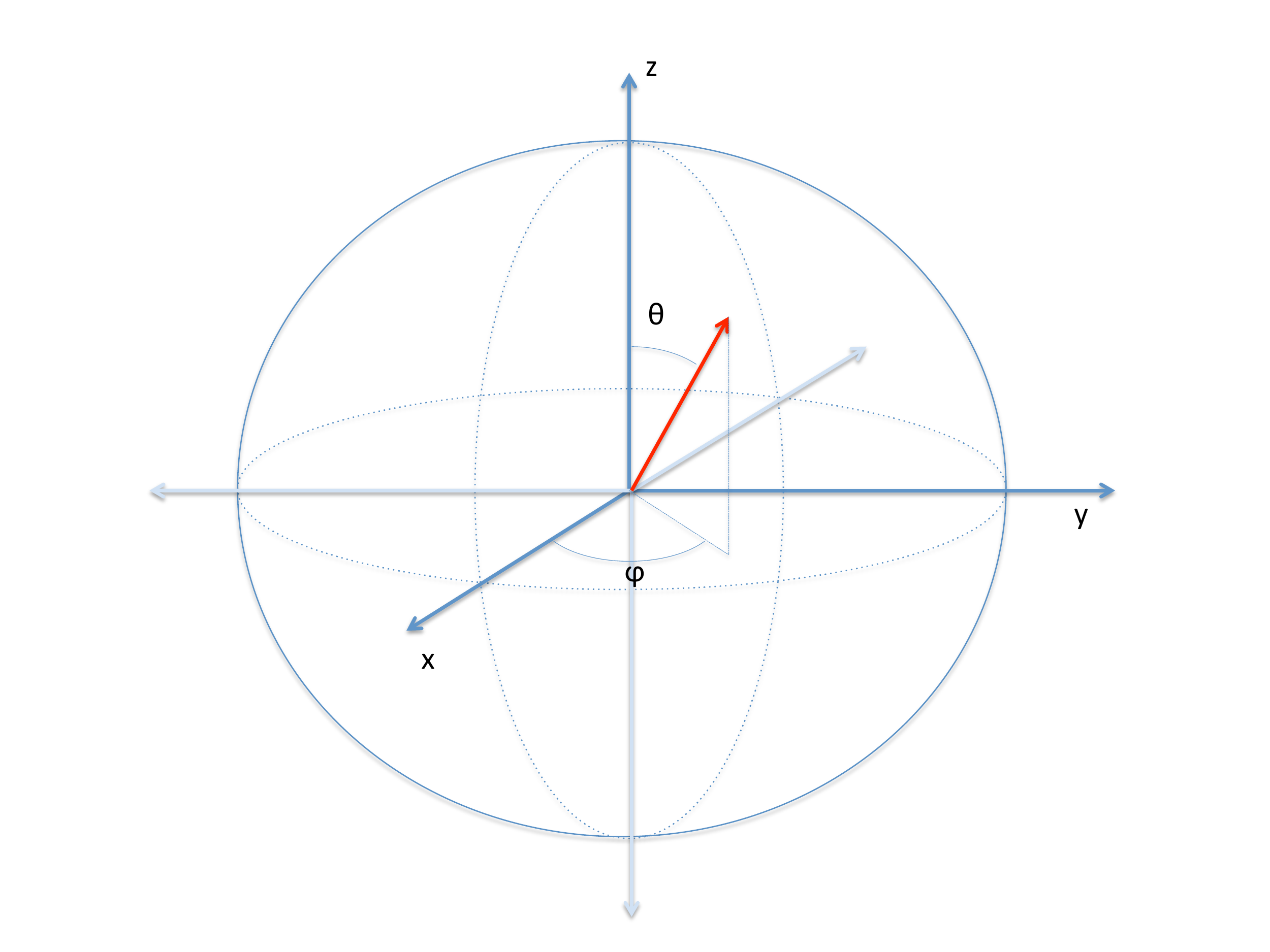}}}}
\caption{The Bloch sphere} 
\label{fig:B}
\end{figure}

Recall its Bloch representation (see e.g. \cite{nc}) as a three dimensional unit vector with spherical polar coordinates $(1, \theta, \varphi)$ (Fig. \ref{fig:B}): 
$$
\psi =(  \sin{\theta} \cos{\varphi},  \sin{\theta} \sin{\varphi}, \cos{\theta})
$$ 
Consider its antipodal vector on the Bloch sphere $\psi^\perp  = R_0(\psi)$,
where $R_0: (x, y, z) \mapsto (-x, -y, -z)$ is the reflection about the origin. It is easy to see that the corresponding quantum state $\ket{\psi^\perp}$ is orthogonal to $\ket{\psi}$.

Thus we can consider the following entangled state on two-qubit system $MA$
$$
\ket{\Omega}_{MT}  = \ket{0}_M \ket{\psi}_T +  \ket{1}_M \ket{\psi^\perp}_T:
$$  
\begin{itemize}
\item The Bloch sphere  of the "type" system $T$ is identified with the Jungian sphere of psychological types.
\item The "mind" system $M$ can be in one of two states, $\ket{0}$ or $\ket{1}$, representing the conscious and unconscious mind, respectively.
\item The system $T$ being in the states $\ket{\psi}$ and $\ket{\psi^\perp}$ corresponds to the conscious and unconscious, respectively.
\item $\ket{\Omega}_{MT}$ is a general maximally entangled state because $\{\ket{\psi}, \ket{\psi^\perp}\}$ is a general orthonormal basis.
\item Thus Jung's theory of types can be formulated as follows: the psychological type of a human being is represented by a unique maximally entangled state.
\end{itemize}

While perhaps elegant, one may object that the quantum representation is a bit of an overkill. In the next section we will see why it is necessary.

\section{Extended Jungian theory and quantum teleportation}
Before we continue it is worth stressing a key assumption made in this work: that Jung was right about pretty much everything. He was a genius when it came to the psyche, and was able to eloquently express his findings using words, symbols and images. Here we simply ask if his insights into the human psyche might be expressed  better and with more precision in the language of mathematics, just like Newton was able to express the laws governing the natural world through the mathematical formalism of classical mechanics.

In his later work  \cite{jung9.1, jung9.2} Jung discovered deeper aspects of the psyche that went beyond the conscious-unconscious dichotomy.
He introduced the following concepts:
\begin{itemize}
\item {\bf The "I".} This is exactly what it sounds like -- the person we consider to be ourselves. In particular the psychological type introduced above refers to this psychological entity.\footnote{To be precise, Jung made a distinction between the "I" and the Shadow, which roughly correspond to the conscious and unconscious. For the purpose of this work we treat them as a single entity.} 
\item {\bf Anima/Animus.} In the psyche of male human there is a female counterpart which he called the Anima. Correspondingly there is a male Animus in the psyche of a woman. The Anima represents the archetypal woman most ideally suited to the personality of the subject in the sense of provoking attraction, fascination and a deep sense of connection at the same time. It may also manifest as the subject taking on the characteristics of the Anima in his own personality, or even becoming  "possessed" by his Anima under certain circumstances.\footnote{Jung himself had visions and dreams related to his Anima throughout his lifetime, and on a concrete level had a lifelong extramarital affair with a woman he identified as being a close representation of his Anima.}
\item {\bf Senex.} Unlike the Anima, the Senex is of the same sex as the subject. The Senex is associated with some perfect version of ourselves, not obtained, however,  through laboriously transforming our ordinary self, but as a  {\it deus ex machina} instant of spontaneous perfection. Every scientist gets a glimpse of this as the solution to a problem comes from nowhere, or an athlete in the moment of "flow". Senex ("Wise Old Man") here refers to glimpses of wisdom, but it could be any other state associated with perfection such as inspiration, beauty or enchantment. Rather than discarding these moments of perfection as statistically insignificant, Jung recognized in them a powerful archetype.\footnote{If the Anima is the archetypal "other" and hence separated from the I {\it in space}, the Senex is separated from the I {\it in time}. }
\item {\bf Self.} The Self is simply the totality of the psyche, and as such subsumes  the I, the Anima and the Senex as integral parts. Jung likened the Self to a mandala, or kaleidoscope-like geometrical pattern. Jung was a man of images, but geometry can only take us so far. An algebraic representation is capable of supporting far greater complexity.
\end{itemize}
We make two remarks:

\begin{itemize}
\item If the Self is to deserve a symbolic representation with nice geometrical properties then there should be a symmetry between the I and the Anima (assuming a male subject), as well a symmetry between the I and the Senex. Furthermore there should be a Senex of the Anima which is the same as the Anima of the Senex. In other words the Self consists of four parts on equal footing: the I, the Anima, the Senex, and the Senex-Anima.
\item If the I is endowed with a psychological type then so should the other three parts of the Self. Moreover the types of the four parts should be related through mathematical transformations. In particular, the Anima should "complement" the I in some way.
\end{itemize}
Those familiar with quantum teleportation \cite{bennett} will already see where this is leading. We claim that the Self can be represented by the following quantum state:

\begin{equation}
  \begin{aligned}
    \ket{\Sigma}_{ASMT} & = \ket{0}_S \ket{0}_A \ket{\Omega}_{MT} \\
     &+ \ket{0}_S \ket{1}_A (I_M \otimes Z_T) \ket{\Omega}_{MT}\\
      & +  \ket{1}_S \ket{0}_A (I_M \otimes X_T) \ket{\Omega}_{MT} \\
      &+ \ket{1}_S \ket{1}_A (I_M \otimes (XZ)_T) \ket{\Omega}_{MT} 
  \end{aligned}
\nonumber
\end{equation}

In a coherent\footnote{Jung viewed the Self as having an identity of its own, beyond the sum of its parts. We thus chose to keep the state  
$\ket{\Sigma}$ coherent, unlike in traditional quantum teleportation.} version of quantum teleportation this is the state of the system before the measurement of the $AS$ system is performed and the appropriate Pauli rotations applied. 
The psychological interpretation of the state $\ket{\Sigma}$ is as follows:
\begin{itemize}
\item The joint computational bases of the "Anima" $A$ and "Senex" $S$ systems distinguishes the four parts of the Self: the I ($\ket{0}_S \ket{0}_A$), the Anima ($\ket{0}_S \ket{1}_A$), the Senex ($\ket{1}_S \ket{0}_A$), and the Senex-Anima ($\ket{1}_S \ket{1}_A$).
\item The system $MT$ contains the corresponding psychological type: if $\ket{\Omega}_{MT}$ is the psychological type of the I, then the type of the Anima is related to it through a Pauli $Z$ rotation, the type of the Senex through a Pauli $X$ rotation, and the Senex-Anima through the composite $XZ$ rotation. 
\end{itemize}

We can now easily work out the psychological types of the four parts of the Self in relation to that of the I. The effect of the Pauli operators on the Bloch sphere is:
\begin{equation}
  \begin{aligned}
Z \ket{\psi} & =  \ket{R_z(\psi)}\\
X \ket{\psi} & =  \ket{R_x(\psi)} \\
XZ \ket{\psi} & =  \ket{R_y(\psi)} 
   \end{aligned}
\nonumber
\end{equation}
where the reflections about the three axes of the Bloch sphere are defined as
\begin{equation}
  \begin{aligned}
R_x: (x, y, z) & \mapsto (x, -y, -z)\\
R_y: (x, y, z) & \mapsto (-x, y, -z)\\
R_z: (x, y, z) & \mapsto (-x, -y, z)
   \end{aligned}
\nonumber
\end{equation}
In terms of the octant approximation this translates into Table \ref{tab:rel}.

\begin{table}[]
\centering
\begin{tabular}{|c|c|c|c|}
\hline
I  & Anima & Senex & Senex-Anima \\ \hline
0i & 3i    & 2e    & 1e          \\ \hline
0e & 3e    & 2i    & 1i          \\ \hline
1i & 2i    & 3e    & 0e          \\ \hline
1e & 2e    & 3i    & 0i          \\ \hline
2i & 1i    & 0e    & 3e          \\ \hline
2e & 1e    & 0i    & 3i          \\ \hline
3i & 0i    & 1e    & 2e          \\ \hline
3e & 0e    & 1i    & 2i          \\ \hline
\end{tabular}
\caption{The psychological types of the four parts of the Self in relation to the psychological type of the I.} 
\label{tab:rel}
\end{table}
 
It is interesting to see how, as one integrates the deeper levels of the Self, one gains access to additional cognitive functions. This explains how one can be of a certain type, yet is able to have some level of understanding of and kinship with those of a different type.
Let us take as an example a male of the 2e type.
\begin{itemize}
\item If he only follows his conscious functions and suppresses the unconscious, he has access to two functions: Te and Ne.
\item As he integrates the unconscious two more functions are included: Fi and Si, even though he is not able to use them in a focussed, conscious way.
\item By the time the Anima (conscious and unconscious part) is integrated, the "missing" functions are included: Fe, Se in the conscious and Ti and Ni in the unconscious. We have collected all eight, but four of them are unconscious. 
\item If one includes the Senex and Senex-Anima, we obtain a "double covering" of the functions: one from the point of view of the conscious and one from the point of view of the unconscious mind. 
\end{itemize}
In Appendix \ref{B} we illustrate the theory with celebrity examples. 
 
\section{Discussion}
Psychology and physics do not enjoy the same credibility. This is because psychology is not expressed in mathematical terms, hence cannot make precise predictions, and hence these predictions cannot be verified. Perhaps  this may now change, as for the first time a  compelling mathematical structure has emerged in the context of psychology.

How does one come to accept a new scientific theory or model? Let us take the example of Einstein's general relativity: 
\begin{enumerate}
\item The model related a new field (theory of gravity) to an existing field (differential geometry) through a simple identification ("matter curves space-time"). There was no a priori reason to think that gravity and geometry should be related, but as soon as the identification was made the theory followed.
\item The model was beautiful and succinct. There is something about beauty that makes it true.
\item The model included Newtonian gravity as a special case.
\item Only later was the model verified experimentally. It could make predictions in the regime where Newtonian gravity was no longer accurate.
\end{enumerate}
Let us see if we can tick these boxes for our extended theory of types.
\begin{enumerate}
\item Psychology and quantum information are clearly pretty different things. Yet Jung's  work on both psychological types and the structure of the Self is all about dualities (conscious-unconscious, intuition-sensation, male-female, etc.), and dualities are bits of information. So the identification here is to say: "Why not quantum information?" And quantum teleportation is arguably the most fundamental protocol of quantum information theory, connecting quantum information, classical information and quantum entanglement.
\item The model presented here owes its aesthetic properties to its equivalence with quantum teleportation.
\item Jung's theory of types is an approximation which only takes the I into consideration.
\item Experimental verification is the remaining point. Over the past five years we have observed hundreds of private individuals as well as public figures, and have no doubt concerning the validity of the theory\footnote{In particular, the choice to associate the Anima and Senex transformations with the $Z$ and $X$ operators, respectively, is in line with our observations}.  What we are lacking is an objective measure. We believe that standardized tests like the Myers-Briggs Type indicator \cite{mbti} are not very accurate. A better approach would be to conduct an in-depth analysis of a sample of patients by appropriately trained psychologists. Apart from ascertaining the psychological type of the person, it is necessary to probe the Anima and observe her psychological type. This is not as hard as it seems because the Anima always finds a way to come out. A much more difficult matter is the Senex. However if one could demonstrate with certainty that the Pauli $Z$ operator is associated with the Anima, this will render the rest of the theory very plausible.  
\end{enumerate}
Jung believed that not all coincidences were actual coincidences. He coined the term \emph{synchronicity} for these "meaningful" coincidences that happened for deeper reasons \cite{jung8}.
Perhaps it was such an act of synchronicity that brought together Jung and quantum pioneer Wolfgang Pauli. Pauli was one of Jung's celebrity patients. His dreams and visions play a central role in one of Jung's books \cite{jung12}. They also collaborated on various ideas related to  symmetry, complementarity and a theory of the Self. Could it be that they were after something like this?

\appendix
\section{Appendix: The eight cognitive functions}
\label{A}
The eight cognitive functions are obtained by combining the four elementary functions, intuition (N), sensation (S), thinking (T) and feeling (F), with the two possible attitudes of introversion (i) and extroversion (e):

\begin{enumerate} 	
\item Extroverted intuition (Ne) is about active perception of the outer world: seeing opportunities in the external world, potential, manoeuvring between obstacles. It is associated with vitality and life energy.
\item Introverted intuition (Ni) is about active perception of the inner world: having panoramic vision on an abstract level, flashes of clarity and sudden understanding. It is associated with "spirit" (as opposed to "matter").
\item	Extroverted sensation (Se) is about passive perception of the outer world: tactile, bodily sensations, the texture of a situation. It is associated with "matter" (as opposed to "spirit").
\item 	Introverted sensation (Si) is about passive perception of the inner world: form as images, symbols, dreams. It is associated with an enchantment-like state.
\item	Extroverted thinking (Te) is about objective judgment of the outer world: facts, information, 
comparisons. The duality is between success and failure. It is associated with goal-oriented action.
\item 	Introverted thinking (Ti) is about objective judgment of the inner world: the essence of ideas, analogical thought, theory. It is associated with "truth", or rather the duality between true and false.
\item 	Extroverted feeling (Fe) is about subjective judgment of the outer world: the duality is between like and dislike. It is associated with "beauty".
\item	Introverted feeling (Fi) is about subjective judgment of the inner world; it is the basis for emotions and particularly associated with "love".

\end{enumerate}

\section{Appendix: Celebrity examples}
\label{B}

Table \ref{tab:celeb} is an attempt to illustrate the theory with a male and female example of each of the eight psychological types, chosen among well known actors and musicians. Combined with Table \ref{tab:rel} one can get a feel for what the Self of a person could look like. For example, a Tom Cruise type has a Nicole Kidman type as his Anima, a George Harrison type as his Senex and a Natalie Portman type as his Senex-Anima.

\begin{table}[]
\centering
\begin{tabular}{|l|l|l|}
\hline
Type & Male  example                & Female   example     \\ \hline
0i   & Johnny Depp            & Joni Mitchell \\ \hline
0e   & Brad Pitt              & Nicole Kidman \\ \hline
1i   & George Harrison          &    Julianne Moore           \\ \hline
1e   & Arnold Schwarzenegger &   Katharine Hepburn            \\ \hline
2i   & Sting                  & Natalie Portman     \\ \hline
2e   &   Mick Jagger      &   Madonna           \\ \hline
3i   & John Lennon           & Barbra Streisand   \\ \hline
3e   & Tom Cruise             & Tina Turner   \\ \hline
\end{tabular}
\caption{Sample celebrity representatives of the eight psychological types} 
\label{tab:celeb}
\end{table}

\end{document}